\documentclass{aa} \usepackage{psfig}

\begin{document}
\def\ntilde{\hbox{\rm n}}
\def\vv{\hbox{\bf v}}
\def\gvec{\hbox{\bf g }}
\def\rvec{\hbox{\bf r }}
\def\svec{\hbox{\bf s }}
\def\vvec{\hbox{\bf v }}


\title{Modeling Non-Confined Coronal Flares: Dynamics and X-Ray Diagnostics}
\author{F. Reale\inst{1} \and F. Bocchino\inst{2} \and G. Peres\inst{1}}
\institute{Dipartimento di Scienze Fisiche \& Astronomiche, Sezione di
Astronomia, Universit\`a di Palermo, Piazza del Parlamento 1, 
I-90134 Palermo, Italy 
\and Osservatorio Astronomico di Palermo G.S. Vaiana,
Piazza del Parlamento 1, I-90134 Palermo, Italy} 

\date{Received /Accepted}
\offprints{F. Reale, reale@oapa.astropa.it}

\abstract{Long-lasting, intense, stellar X--ray flares may approach
conditions of breaking magnetic confinement and evolving in open space.
In the perspective of searching for possible tracers of
non-confinement, we explore this hypothesis with hydrodynamic
simulations of flares occurring in a non-confined corona: model flares
are triggered by a transient impulsive heating injected in a
plane-parallel stratified corona. The plasma evolution is described by
means of a numerical 2-D model in cylindrical geometry $R,Z$. We
explore the space of fundamental parameters. As a reference model, we
consider a flare triggered by a heating pulse of 10 erg cm$^{-3}$
s$^{-1}$ lasting 150 s and released in a region $\sim 10^9$ cm wide and
at a height $\sim 2 \times 10^9$ cm from the base of the stellar
surface.  The pressure at the base of the corona of the unperturbed
atmosphere is 0.1 dyne cm$^{-2}$.  The heating would cause a 20 MK
flare if delivered in a 40000 km long closed loop.  The modeled plasma
evolution in the heating phase involves the propagation of a 10 MK
conduction front and the evaporation of a shocked bow density front
upwards from the chromosphere. As the heating is switched off, the
temperature drops in few seconds while the density front still
propagates, expanding, and gradually weakening. This kind of evolution
is shared by other simulations with different coronal initial pressure,
and location, duration and intensity of the heating.  The X-ray
emission, spectra and light curves at the ASCA/SIS focal plan, and in
two intense X--ray lines (Mg XI at 9.169~\AA~and Fe XXI at
128.752~\AA), have been synthesized from the models. The results are
discussed and compared to features of confined events, and scaling laws
are derived.  The light curves invariably show a very rapid rise, a
constant phase as long as the constant heating is on, and then a very
fast decay, on time scales of few seconds, followed by a more gradual
one (few minutes). We show that this evolution of the emission, and
especially the fast decay, together with other potentially observable
effects, are intrinsic to the assumption of non--confinement. Their
lack indicates that observed long--lasting stellar X--ray flares should
involve plasma strongly confined by magnetic fields.  
}
\maketitle

\section{Introduction}
\label{sec:intro}

Stellar coronae are not spatially resolved with present day
observations. It is widely accepted that they share many of the basic
mechanisms of the solar corona, the only resolved corona that we know.
Decades of X-ray observations have shown that both its structuring and
heating are dominated by the ambient magnetic field.  Evidence has been
found, such as the presence of coronal plasma at temperatures of
several million degrees (e.g. Linsky 1981, Schmitt et al. 1990), that
some form of confinement of X-ray bright plasma should be at work on
stars, probably associated with intense magnetic fields.  It is then
natural to wonder what is the morphology of the confining structures in
a stellar corona. On the Sun, plasma is typically confined in magnetic
arches (the loops), and both the geometry and luminosity of many of
them are steady for relatively long times.

If stellar coronae are similar to the solar one, then structures similar 
to solar loops should be found on stars. Several approaches are possible 
to determine the dimensions of coronal structures. An eclipse on Algol 
has allowed to find the volume responsible of a flare (Schmitt \& Favata 
1999). Loop fitting has been used to constrain the average linear 
dimensions of stellar coronal loops (e.g. Maggio \& Peres 1997). Another 
source of information on the characteristic of stellar coronal loops is 
the evolution of X-ray flares. In particular, basic mechanisms of plasma 
cooling are radiation and thermal conduction from the corona downwards 
to the cooler chromosphere. Linking the plasma characteristic cooling 
times to the observed flare decay time allows to infer the linear size 
of the cooling region. This property has been largely exploited in the 
past to estimate the size of stellar flaring loops (e.g. Haisch 1983,
van den Oord et 
al. 1988, van den Oord \& Mewe 1989, Reale et al.  1988, Reale et al. 
1997, Favata \& Schmitt 1999, see Reale 2001 for a review).  

One implication of the simple scaling from cooling times is that slower 
and slower decays, associated with long-lasting flares, correspond to 
longer and longer loops. As a matter of fact, stellar X-ray flares are 
typically observed to last much longer than average solar X-ray flares, up to 
several days whereas the duration of solar flares ranges from few 
minutes to several hours. In spite of the obvious observational bias 
toward detecting long events, the evidence of long flares has been taken 
as strong indication of very large flaring regions, in particular 
comparable to, or even larger than, the stellar radius (e.g.  
Graffagnino et al. 1995, Favata \& Schmitt 1999, Tsuboi et al. 2000, 
Favata 2001).  On the Sun, large scale and long-duration events 
typically occur in complex coronal structures, in which the magnetic
field undergoes major rearrangements and several loop structures are
progressively involved (e.g. two-ribbon flares): the large amounts of
energy released lead to large plasma pressure which may exceed the
magnetic one and cause the end of confinement.  The question now is:
may plasma in long-lasting and intense stellar flares break the
magnetic cage, and erupting, be free to move and evolve in open space?

This work addresses this question, by modeling flares occurring in a
totally non-confined stratified corona. Although we do not pretend this
to be an entirely realistic scenario, it can be considered as an
extreme limit of situations in which the plasma dynamics dominates over
the confining magnetic field and therefore governs the X-ray luminosity
evolution.

Indeed, some solar and stellar flares show phenomena that indicate that
at least some of the plasma is not confined. In the solar case, coronal
mass ejections (CMEs) following flares are not confined. In the stellar
case VLBI measurements of nonthermal particles propagating away from
the star are observed. A good example is a flare in UX Ari observed by
Mutel et al (1985). White and Franciosini (1995) also show that the
relativistic plasma responsible for the gyrosynchrotron radio emission
in stellar comes from expanding plasma.

Here we focus on studying the X-ray emitting plasma. Our scope is to
explore: to what point actual stellar flares can approximate
free-expanding flaring plasma, if there are distinctive signatures in
X-ray emission (obtained from the modeling) which characterize
unconfined flares vs confined ones and therefore allow us to
discriminate the confining role of the magnetic field, and if observed
stellar flares present such signatures.

The study that we illustrate here presents also interesting results on
the theoretical point of view: we describe the evolution of the
non-confined plasma in a stratified coronal atmosphere, i.e. which is
hot {\it ab initio} and where thermal conduction is fully efficient.
This aspect makes this work quite different from typical models of
bursts, like supernovae, which instead propagate in non-conducting and
cooler media.

Our approach is to set up an initial stratified atmosphere, similar to
a confined one, including chromosphere, transition region and corona,
and where the plasma is not confined to move along one spatial
direction, and then to impart a localized energy impulse. Plasma
dynamics and heat conduction can occur isotropically. The evolution of
such system needs to be described by a numerical 2-D hydrodynamic
model, with enhanced spatial resolution in the low corona and in the
transition region and including plasma thermal conduction; therefore it
is also a demanding numerical problem. In order to compare our results
with observations and to obtain diagnostical feedbacks, we synthesize
from our model results the expected evolution of the integrated X-ray
flare luminosity, i.e. by simulating a flaring unresolved source, such
as a star. To have a realistic scenario, representative of real flare
observations, we have chosen to synthesize the emission in a wide X-ray
band such as at the focal plane of the ASCA/SIS (sensitive
in the range, approximately, 0.3 - 10 keV), and in two intense X-ray
lines. Similar results can be expected for other wide band CCD
instruments such as Chandra/ACIS and XMM/EPIC, while the lines may be
detectable by instruments with good spectral resolution such as Chandra and
XMM-Newton grating detectors.

In Sect.~\ref{sec:model} we describe our modeling approach in detail,
in Sect.~\ref{sec:simul}, the simulations performed and the results
obtained are presented, in Sect.~\ref{sec:discuss} the results are
discussed and in Sect.~\ref{sec:conclu} conclusions are drawn.

\section{Modeling}
\label{sec:model}
\subsection{The Model Set Up}

The general concept of our modeling non-confined coronal flares is
similar to modeling confined flares (e.g.  Peres et al. 1982):  the
flare is triggered by an impulsive heating injected in a stratified
plane-parallel static corona.  The basic difference is that the
unperturbed atmosphere is not a semicircular coronal loop and that the
heat transport, as well as any plasma motion, is allowed in any
direction, instead of being limited only along magnetic field lines
parallel to the loop flux tube. The evolution of a heat perturbation
propagating in any direction in a stratified atmosphere can be properly
described with a 2-D cylindrical geometry ($R, Z$).  This description
is adequate for our purposes of deriving the global properties of the
plasma evolution and X-ray emission as it would be detected at stellar
distances.

\begin{figure*}
\centerline{\psfig{figure=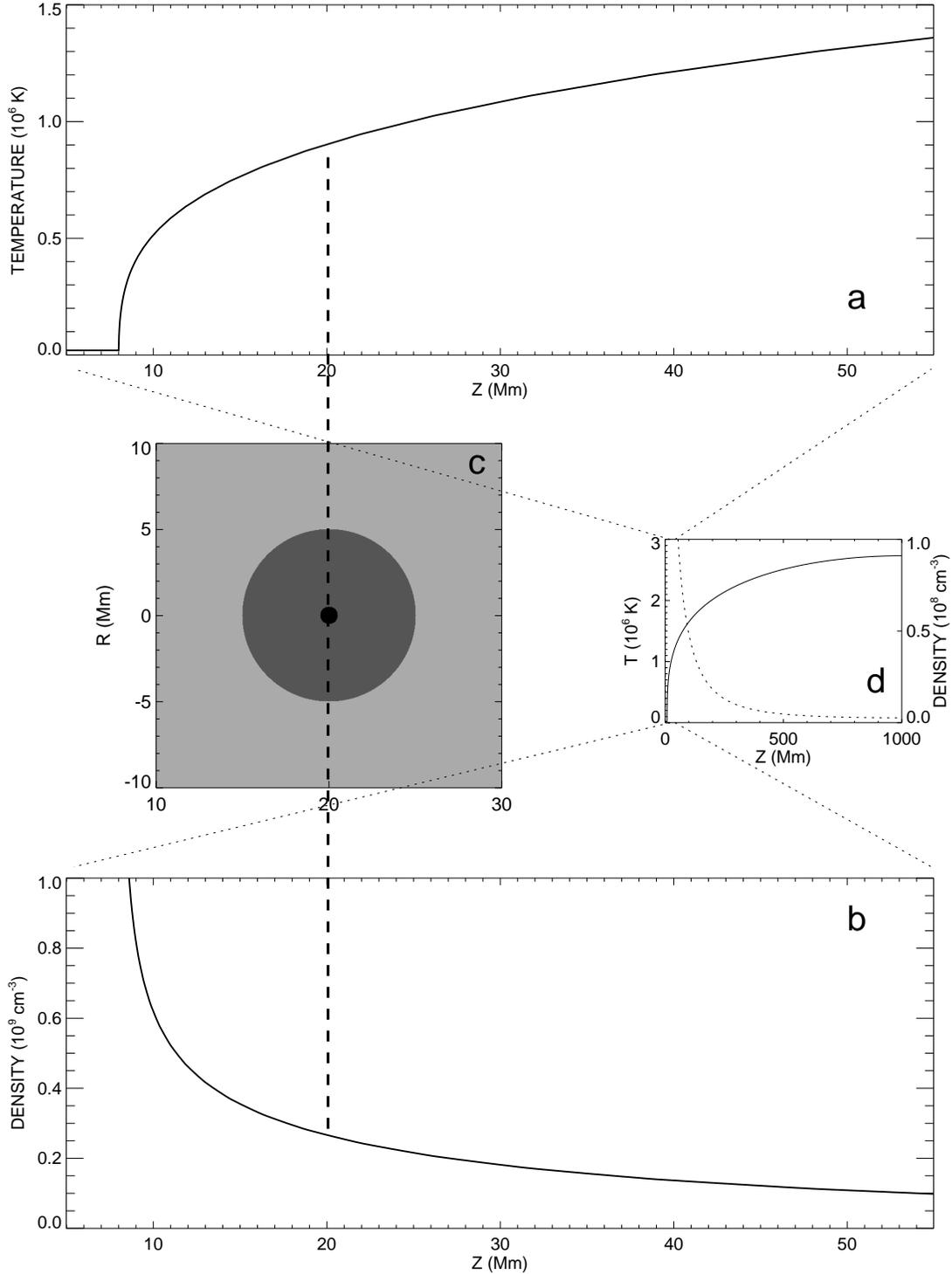,width=15cm}}
\caption[]{Initial configuration of the reference model (cf.  model 1
in Table 1); the geometry is cylindrical with coordinates $r, z$.
Panels (a) and (b) show the temperature and density distributions along
the vertical $z$ axis ($r=0$).  Panel (c) shows the spatial
distribution of the impulsive heating vs radial distance ($R$) and
height ($Z$), where the radius of the dark grey circle is $\sigma_H$
(see Eqs.~\ref{eq:heat},~\ref{eq:heat2}). Notice that the upward vertical
direction is rightwards in panel (c). Panels (a), (b) and (c) show
a limited region of the whole computational domain.  Panel (d) shows
the temperature and density ({\it dashed}) distributions along the
vertical $z$ direction on a much larger scale.
\label{fig:model}}
\end{figure*}

The solar template of confined coronal flares has guided us in
configuring the initial unperturbed atmosphere and in tuning the plasma
and heating parameters.  As shown in Fig.~\ref{fig:model}, the initial
atmosphere consists of a stratified plane-parallel corona extending
vertically from the stellar surface. The limit of our computational
domain is at $10^{11}$ cm from the stellar surface. The corona is
maintained steady at coronal conditions by a uniformly distributed
heating. As our reference starting conditions we have taken an
atmosphere with a pressure at the base of the corona of 0.1 dyne
cm$^{-2}$, typical of coronal quiet regions, which may reasonably
approximate non-confined coronal conditions. In these conditions the
temperature of the upper layers of the non-flaring corona of our
computational domain is below 3 MK. We have made some simulations also
with a base pressure 1 dyne cm$^{-2}$, a denser and hotter ($\le 6$ MK)
initial atmosphere, to check the effect of changing the initial
conditions to rather extreme cases of very active "non-confined"
coronae. The bottom of the corona is matched to a much cooler
chromosphere through a steep transition region. All this setting is
similar to that for modeling confined coronal flares (e.g. Peres et
al.  1982, Cheng et al. 1983, McClymont \& Canfield 1983). The
chromosphere is required as mass reservoir for the flare evolution; our
choice has been to consider an isothermal ($2 \times 10^4$ K) and
non-radiating chromosphere (Cheng et al. 1983). While the detailed
structure of the chromosphere is important for the fine details of the
evaporation and coronal evolution, a coarse chromospheric model like
the one used here suffices to show the general characteristics of the
evolution addressed here.  We have verified that the unperturbed
atmosphere is steady and stable for times much longer than the flare
evolution times considered for this study.

The plasma evolution is described by solving the 2-D plasma hydrodynamic 
equations of conservation of mass, momentum and energy with the geometry 
described above:

\begin{equation}
\frac{\partial n} {\partial t} + \nabla \cdot (n \vvec)
= 0 \,,
\end{equation}

\begin{equation}
\frac{\partial (n \vvec)}{\partial t} + \nabla \cdot (n
\vvec \vvec) = - \frac{1}{m} \nabla p + n \gvec \,,
\end{equation}

\begin{equation}
\frac{\partial u} {\partial t} + \nabla \cdot \left [ (u + p)
\vvec \right]= m n \gvec \cdot \vvec - n^2 P(T) + {\cal H}(\rvec) - 
\nabla F_c \,,
\label{eq:energy}
\end{equation}

\begin{equation}
u = \frac{1}{2} m n v^2 + {\cal E} \,, ~~~
p = (\gamma - 1) {\cal E} \,, ~~~
T = \frac{p}{2 k n} \,, ~~~
F_c = - \kappa T^{5/2} \nabla T
\end{equation}

\noindent
where $n$ particle density per unit volume, $t$ time, $\vvec$ the
plasma bulk velocity, $p$ pressure, $m$ average mass per particle ($m =
2.1 ~ 10^{-24}$ g for solar abundance), $\gvec$ gravity acceleration,
$P(T)$ radiative losses per unit emission measure, ${\cal H}$ heating
per unit volume (uniform and constant in transition region), $\cal E$
is the internal energy density, $T$ plasma temperature, $F_c$ heat
flux, $\kappa$ plasma thermal conductivity and $\gamma = 5/3$.  The
flare is assumed to occur on a solar-like star, with solar surface
gravity and radius.

Plasma thermal conduction is isotropic.  Radiative losses are those of
Raymond and Smith (1977) but are set to zero for $T \approx 2\times
10^4$ K, i.e. for chromospheric plasma. The heating term generally
consists of a steady heating term which maintains the unperturbed
atmosphere in thermal equilibrium and of a transient term, which
triggers the flare. In our simulations, the steady heating has been set
to zero both in the chromosphere and elsewhere, since the transient
heating is switched on, in order to describe the free plasma cooling
with no interference from any heating source.

The transient heating term has been assumed as a separable function of 
space and time:

\begin{equation}
H(Z,R,t) = H_0 f(Z,R) g(t)
\label{eq:heat}
\end{equation}
The spatial distribution of the heating is a 2-D circular Gaussian:
\begin{equation}
f(Z,R) = \exp \left[ \frac{(Z-Z_H)^2+(R-R_H)^2}{2 \sigma_H^2} \right]
\label{eq:heat2}
\end{equation}

As for the temporal evolution g(t) the impulsive heating has been
assumed to be switched on at t=0 ($g(0)=0$), kept constant ($g(t)=1$)
for a given time lapse $0 \leq t \leq t_H$ and then switched off $g(t >
t_H ) = 0$.

\subsection{The Code}

The conservation equations shown above are solved numerically on a
fixed geometrical grid. The time integration is made with ``time
splitting" in two separate phases:  a) the proper hydrodynamic phase,
which includes the mass and momentum equations and the advective terms
of the energy equation; b) the thermal conduction phase, which includes
also radiation and heating processes. The hydrodynamic section is based
on an explicit Flux Corrected Transport (FCT) numerical scheme,
improved for efficiency and accuracy in large-scale smooth regions
(Reale et al. 1990). The highly non-linear thermal conduction is
treated with an Alternating Direction Implicit (ADI) scheme (Reale
1995), which warrants unconditional stability.  The two phases are
solved independently according to a time-splitting approach, i.e. using
different integration time steps, which depend on the characteristic
times of the physical effects involved in each section and on the
stability criteria of the respective numerical schemes, i.e. FCT and
ADI.

As mentioned above the geometrical domain is 2-D in cylindrical
coordinates ($R, Z$). The grid does not change during a simulation and
the grid cells are rectangular: the spacing $dZ$ along the $Z$ direction is
uniform in the chromosphere ($dZ = 2 \times 10^6$ cm) and then
monotonically increases with a geometric progression up to $dZ \sim 2 \times
10^9$ cm at $Z \approx 10^{11}$ cm; this spacing warrants a stable and
steady atmosphere in equilibrium conditions; the spacing along $R$ is
uniform as far as $7 \times 10^{10}$ cm from the central axis and then
monotonically increases as far as $\sim 10^{11}$ cm. The spacing along
$Z$ and $R$ has been chosen so as to have a relatively high resolution
and a regular grid in the internal domain region, where the energy is
released and the flare mostly occurs, and at the same time, a very
large domain (Bedogni \& Woodward 1990), so that the boundary
conditions at the far extremes become unimportant for the problem under
analysis. The space cells along $Z$ are 800 for all simulations
performed. Along $R$ most simulations has been done with 200 grid cells
for the internal uniform region, plus 28 cells of non-uniform expanding
grid; some simulations have been done with 100 and 400 grid cells for
the internal region.

Reflecting boundary conditions have been set for the central $Z$ ($R=0$) 
axis. Fixed density and temperature have been set constant on the lower 
$R$ axis ($Z=0$), and with zero gradient on the upper one. The $Z$ 
component of the velocity is set to zero at both $Z$ boundaries, the 
$R$ component with zero gradient. At the far extreme of the $R$ domain 
zero gradient conditions have been set for all the relevant quantities.

The code has been parallelized on the basis of a geometric domain
decomposition and using the High Performance Fortran (HPF).  In the
Alternating Direction Implicit (ADI) scheme, the equation is integrated
at each time step in two half steps.  At each half-step the numerical
integration advances in one of the two spatial directions.  The
integration scheme at each half-step is implicit: the solution at one
grid-point is derived from the solution at the same time step at an
adjacent point. The data dependencies prevent direct parallelization of
the nested DO loops, as in the case of the FCT section:  we have
adopted a parallelization solution which minimizes the interprocessor
communications and the sequential operations, without changing the
numerical algorithm.  Simulations have been executed on the Cray T3E at
CINECA (Bologna, Italy), on a Compaq server and on a Compaq HPC Cluster
at Osservatorio Astronomico di Palermo (Italy).

\section{The Simulations}
\label{sec:simul}

\subsection{The Parameter Space and The Reference Flare}
\label{sec:parref}

Table~\ref{tab:sim} shows the parameters and some results of the
simulations of non-confined flares presented here (plus one model of
confined flare for comparison). For each simulation, there are the
basic simulation parameters, i.e. the pressure at the base of the
initial unperturbed corona, and the parameters of the heating release
(see Eqs.~\ref{eq:heat} and \ref{eq:heat2}), i.e. the heating rate per
unit volume $H_0$, the height $Z_H$ of the heating maximum, the
Gaussian width $\sigma_H$, the total rate, and the duration of the
heating (Eq.~\ref{eq:heat}), and some relevant results, i.e. the plasma
maximum temperature, the maximum density at the position of the maximum
temperature, the maximum velocity throughout the computed flare
evolution, and, finally, the light curve decay time after the initial
very fast decay phase (see Sect.~\ref{sec:emiss}).

Our simulation strategy is to set up and compute a reference model case
and then perform other calculations by changing the parameters relevant
for our study one by one.  Our reference configuration has a coronal
base pressure $p_0 = 0.1$ dyne cm$^{-2}$, which is a value typical of
quiet coronal regions or even of open coronal structures on the Sun.
Since we suppose that higher pressures may occur on active stars,
values up to  $p_0 = 1$ dyne cm$^{-2}$ have also been considered.

In our reference simulation the flare is triggered by a heating pulse with
maximum rate rate per unit volume $H_0 = 10$ erg cm$^{-3}$ s$^{-1}$ 
and width of the Gaussian $\sigma_H = 5 \times 10^8$ cm centered
on the $Z$ axis ($R_H = 0$) at a height of $Z_H = 2 \times
10^9$ cm above the base of the atmosphere, well above the transition
region (Eq.~\ref{eq:heat2}). Similar heating rate, width and position
would, for instance, ignite a 20 MK confined solar flare at the top of
an active region loop (Peres et al.  1987), which may be taken for
comparison to the results presented here. As shown in
Table~\ref{tab:sim}, other simulations have been performed with higher
values of the rate $H_0$ up to 100 erg cm$^{-3}$ s$^{-1}$, in order to
explore the trend toward more intense events such as those typical of
active stars. We have also considered heat deposited higher in the
corona, i.e. more distant from the stellar surface, and, at the same
time, over a more extended volume, while either maintaining the same
total heating rate or doubling it.  In order to obtain a significant
brightening over the background atmosphere, in all simulations with
higher base coronal pressure, the heating rate $H_0$ is higher than the
reference value, since the atmosphere is brighter and also more efficient in
radiating the additional heat, this making any smaller heating barely
effective.

For most simulations, the duration of the constant heating (the time in
which $g(t) = 1$) is 150 s, but a case of much longer heating ($t_H =
600$ s) has also been explored.

The reference simulation (named {\it ref} in Table~\ref{tab:sim}) will
be described in more detail in the following.  With the total heating
rate of $\sim 2 \times 10^{28}$ erg s$^{-1}$, the total heat released
is therefore $\sim 3 \times 10^{30}$ erg. The evolution is computed for
a total time of 800 s, i.e. the decay is studied for 650 s.  The grid
is $800 \times 228$ grid cells.

For comparison with the models of flares in non-confined atmospheres,
Table~\ref{tab:sim} shows also the parameters and results of a model of
flare in a {\it confined} coronal loop, labelled with {\it Conf}. In
particular, the parameters are those of a flare in an active region
loop of half-length $2 \times 10^9$ cm and initial pressure 6 dyne
cm$^{-2}$ (Peres et al. 1982, Peres et al. 1987). The heating duration,
spatial extent and intensity per unit volume have been chosen identical to those
of the reference model. A loop cross-section area of $2.5 \times 10^{17}$
cm$^2$ has been assumed, corresponding to a loop aspect, i.e. radius divided by
half-length, of 0.1.

\begin{table*}
\begin{center}
\caption[]{\label{tab:sim} Parameters of simulated non-confined flares $+$
one confined flare\\}
\begin{tabular}{lcccccccccc}
\hline
Name&p$_0^a$&H$_0^b$&Z$_H^c$&$\sigma_H^d$&H$_{tot}^e$&	$t_H^f$&T$_{max}^g$&
D$_{max}^h$&$V_{max}^i$&$\tau_{dec}^j$ \\
&dyn/cm$^2$& erg cm$^{-3}$ s$^{-1}$&$10^9$ cm&$10^9$ cm&
$10^{28}$erg/s&s&$10^7$ K&$10^9$ cm$^{-3}$&km/s&s \\
\hline
Ref & 0.1 & 10 & 2.0 & 0.5 & 2 & 150 & 1.3 & 0.9 & 800&290 \\
Long & 0.1 & 10 & 2.0 & 0.5 & 2 & 600 &  1.3 & 0.9 & 1000& 380 \\
H30 & 0.1 & 30 & 2.0 & 0.5 & 6 & 150 & 1.7 & 1.2 &1000&  270 \\
H100 & 0.1 & 100 & 2.0 & 0.5 & 20 & 150 & 2.5 & 1.9 & 1200& 320 \\
X5H5 & 0.1 & 5 & 5.0 & 1.0 & 8 & 150 & 1.6 & 0.3 & 700& 180 \\
X5H10 & 0.1 & 10 & 5.0 & 1.0 & 16 & 150 & 2.0 & 0.4 & 900& 210 \\
P03H30 & 0.3 & 30 & 2.0 & 0.5 & 6 & 150 & 1.7 & 1.7 & 1000& 350 \\
P03H100 & 0.3 & 100 & 2.0 & 0.5 & 20 & 150 & 2.5 & 2.6 & 800& 370 \\ 
P1 & 1.0 & 100 & 2.0 & 0.5 & 20 & 150 & 1.9 & 2.8 & 600& 200 \\
\hline
Conf$^k$ & 6 & 10 & 2.0 & 0.5 & 0.2 & 150 & 2.1 & 75 & 360&250$^l$ \\
\hline
\end{tabular}
\end{center}
\noindent
$^a$ Pressure at the base of the corona

\noindent
$^b$ Maximum heating rate per unit volume (see Eq.\ref{eq:heat})

\noindent
$^c$ Height of the center of the heating (see Eq.\ref{eq:heat})

\noindent
$^d$ Width of the heating (see Eq.\ref{eq:heat})

\noindent
$^e$ Total heating rate

\noindent
$^f$ Heating duration (see Eq.\ref{eq:heat})

\noindent
$^g$ Flare maximum temperature

\noindent
$^h$ Maximum density at the position of the temperature maximum

\noindent
$^i$ Maximum velocity

\noindent
$^j$ Light curve (in the ASCA/SIS band) e-folding decay time after the
initial fast decay phase

\noindent
$^k$ Model of a flare confined in a solar active-region loop

\noindent
$^l$ E-folding decay time since the light curve maximum

\end{table*}

\subsection{The plasma evolution of the reference flare}
\label{sec:evol}

\subsubsection{The heating phase}
\label{sec:evol_heat}

\begin{figure*}
\centerline{\psfig{figure=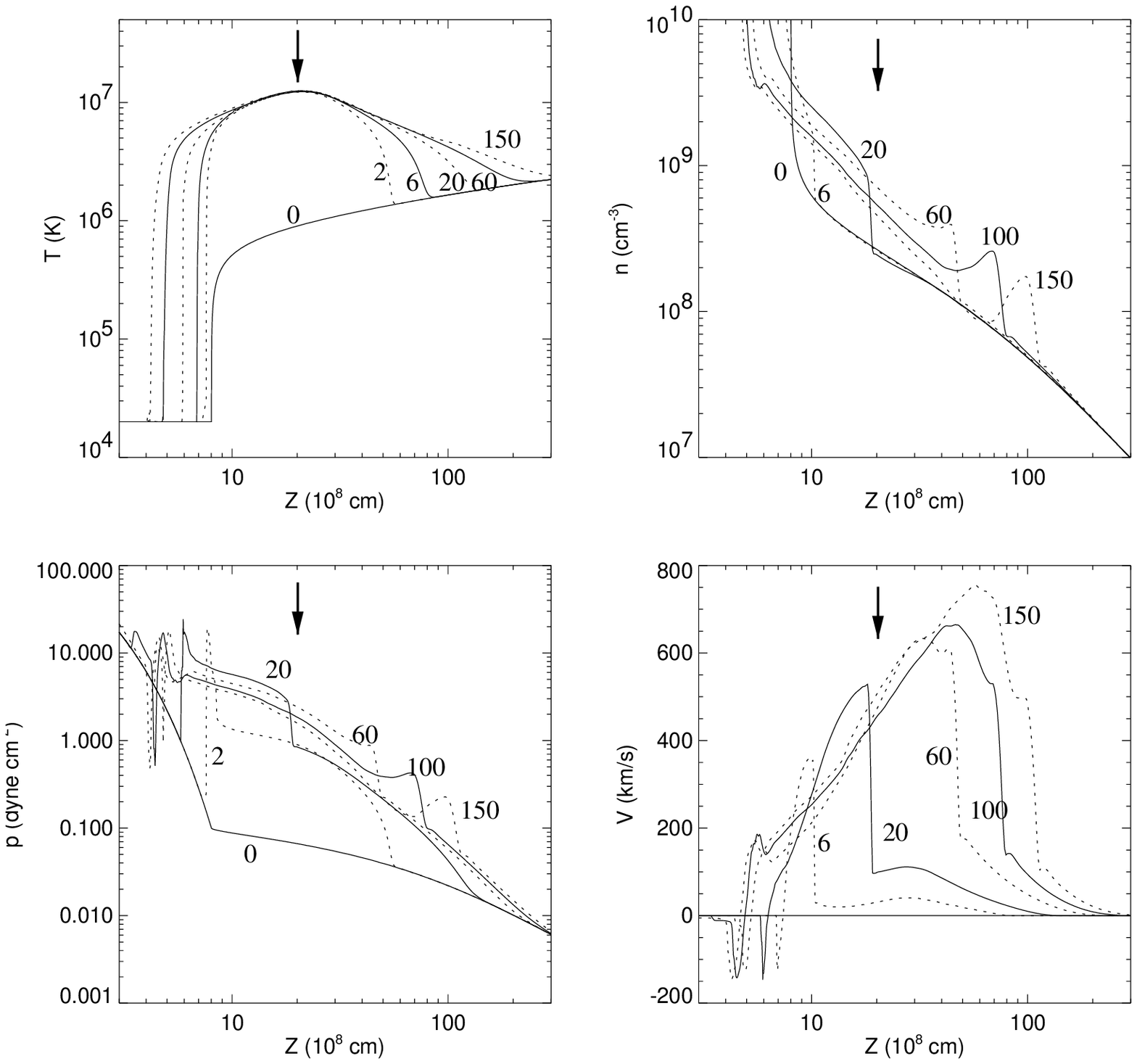,width=14cm}}
\caption[]{Distributions of temperature $T$, density $n$, pressure $p$,
and vertical component of velocity $V=V_Z$, along the central $Z$
axis ($R=0$) and at various times (s) during the heating phase of the
reference simulation of the non-confined flare. The arrows indicate the
position of the center of the heating distribution.
\label{fig:hydro1}}
\end{figure*}

\begin{figure*}
\centerline{\psfig{figure=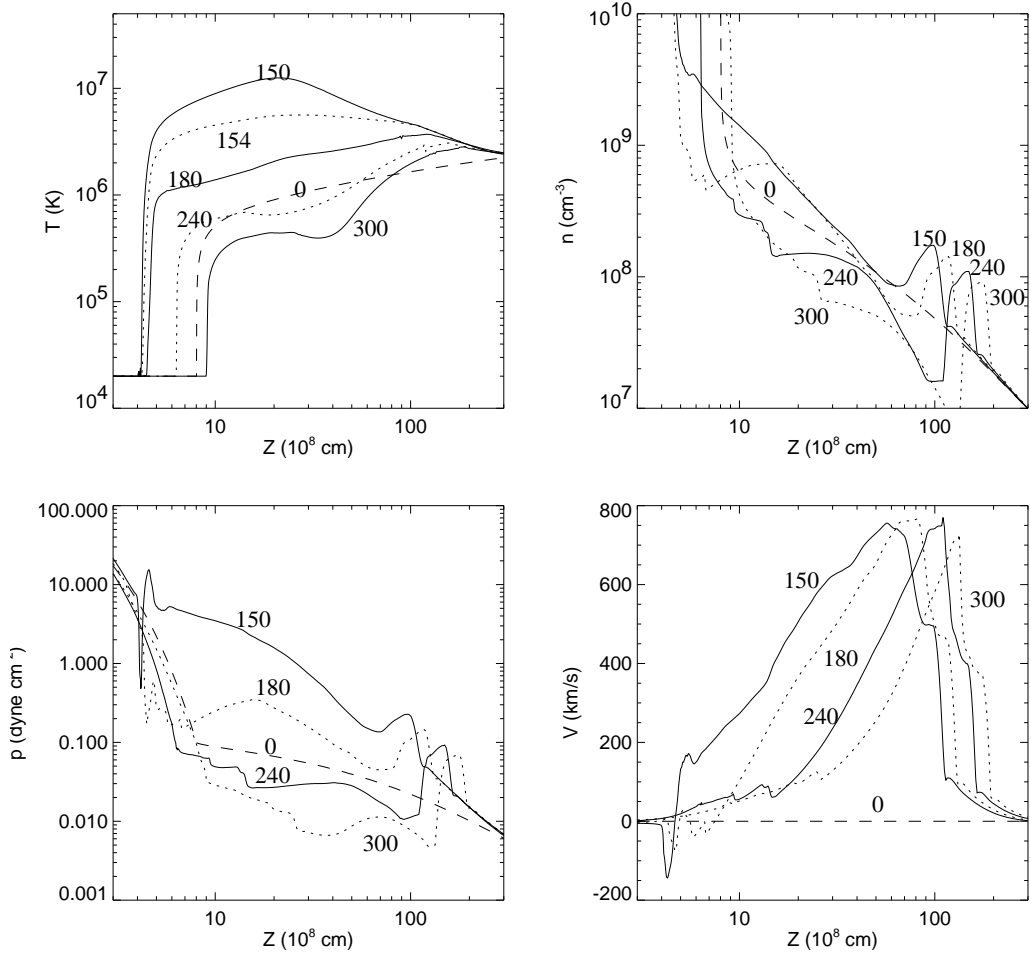,width=14cm}}
\caption[]{As in Fig\ref{fig:hydro1} but taken at times after the heating is
switched off.
\label{fig:hydro2}}
\end{figure*}

\begin{figure*}
\centerline{\psfig{figure=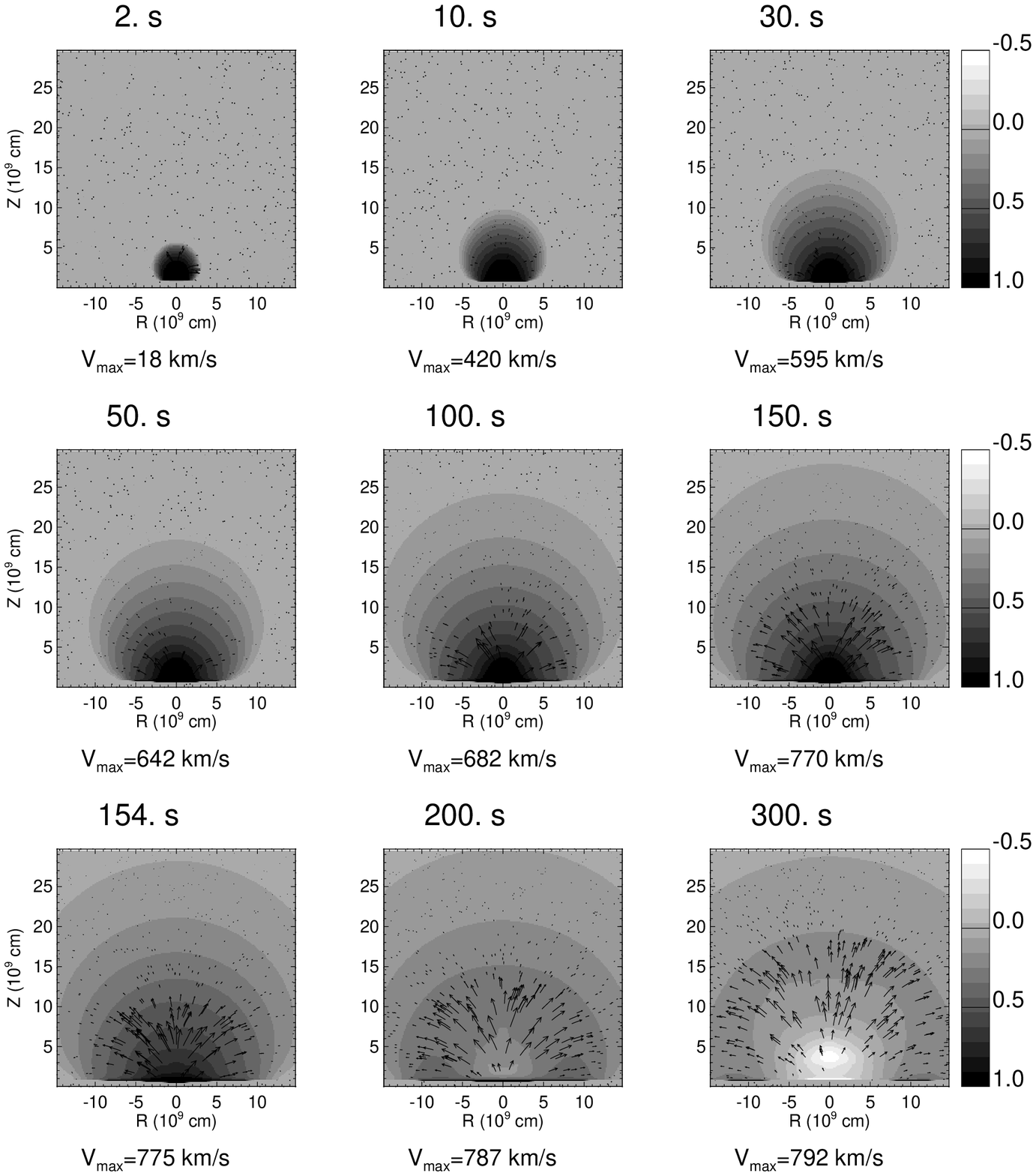,width=14cm}}
\caption[]{Grey scale images of the temperature contrast $T/T_0$, where
$T_0$ is the temperature distribution of the unperturbed atmosphere,
and velocity field ({\it arrows}) at various times during the reference
simulation. It is shown the region around the 
heating deposition.
The grey-scale is logarithmic. The maximum speed
corresponding to the longest arrow is shown below each image.
\label{fig:imtemp}}
\end{figure*}

\begin{figure*}
\centerline{\psfig{figure=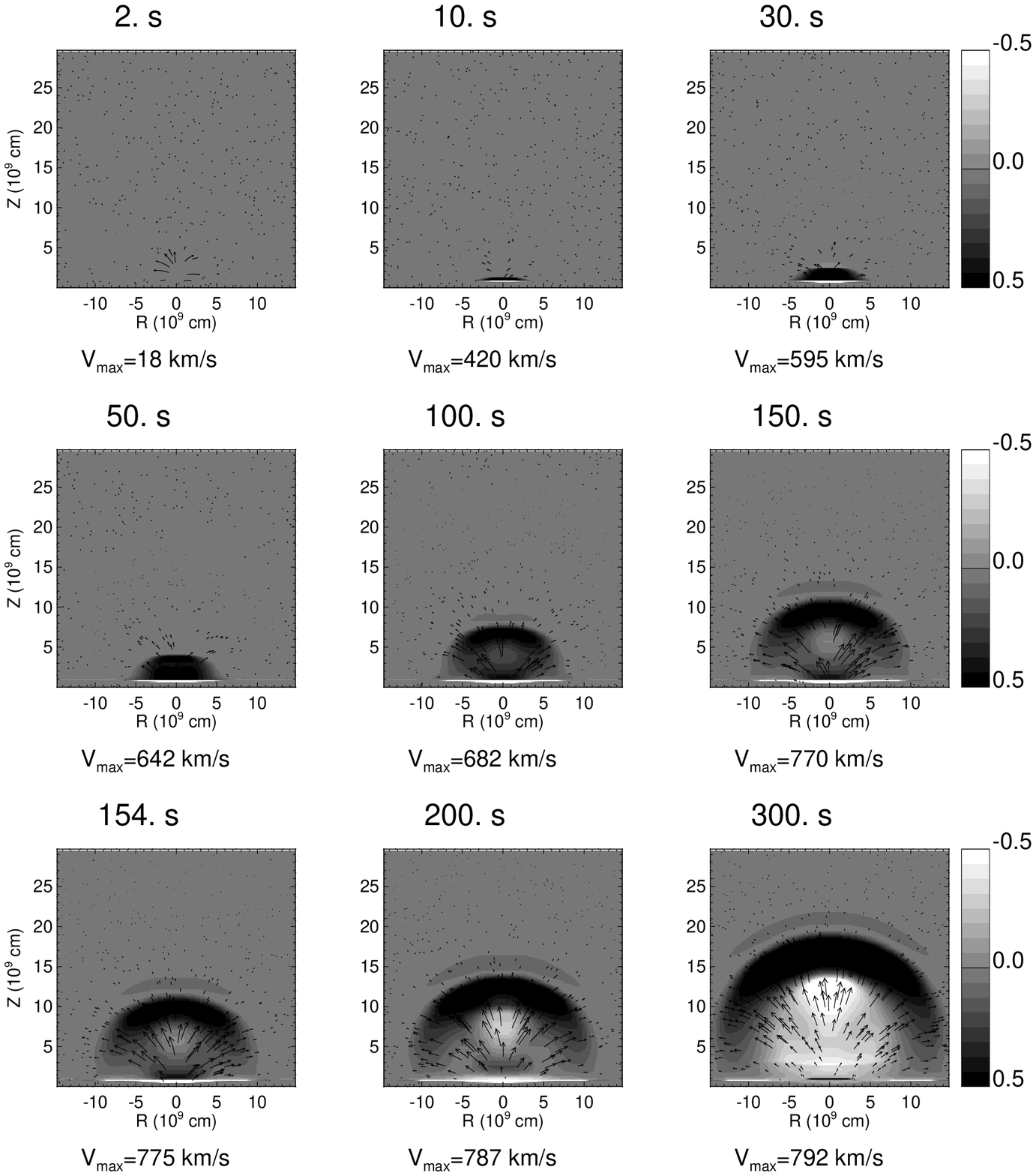,width=14cm}}
\caption[]{As in Fig.\ref{fig:imtemp} but for the density contrast
$n/n_0$, where $n_0$ is the density distribution of the unperturbed atmosphere.
\label{fig:imdens}}
\end{figure*}

Fig.~\ref{fig:hydro1} shows distributions of temperature, density,
pressure and vertical component of velocity along the central Z axis at
various times during the phase in which the heating is switched on in
the reference simulation. Fig.~\ref{fig:hydro2} shows the analogous
results of the evolution after the heating has been switched off.
Figs.~\ref{fig:imtemp} and \ref{fig:imdens} respectively show grey-scale
images of the temperature and density contrasts $T/T_0$ and $n/n_0$,
where $T_0$ and $n_0$ are the temperature and density of the
unperturbed atmosphere at various times during the same simulation,
including both heating and decay phase. Figs.~\ref{fig:hydro1},
\ref{fig:imtemp} and \ref{fig:imdens} clearly show the evolution of the
plasma in the heating phase.

Because of the strong impulsive heating the plasma temperature at the 
center of the heating release on the $Z$ axis increases rapidly (in less 
than 2 s) from 1.5 to 12 MK (and similarly the pressure by a factor 
10) and then remains practically constant there. A thermal conduction 
front develops and propagates spherically and very rapidly in such a thin 
corona, taking $\sim 2$ s to reach the base of the corona, the 
transition region and the chromosphere, and, on the opposite side, a height
$Z \sim 5 \times 10^9$ cm. As the conduction front propagates in
corona, expanding from the heating region, it fades and slows down
significantly. While a height $Z = 10^{10}$ cm is reached by the
temperature change after only $\sim 15$ s, it takes one minute more
to reach $Z = 2 \times 10^{10}$ cm, and at the end of the heating the
conduction front is only slightly higher than that. In the orthogonal
(radial) direction the conduction front is slower, because the
temperature increases upwards, and therefore so does the efficiency of
thermal conduction, while this does not occur in the radial direction;
$R = 10^{10}$ cm is reached only after 50 s since the beginning of the
simulation and at the end of the heating the distance of the conduction
front from the $Z$ axis is $R = 1.5 \times 10^{10}$ cm. To have an idea
of the weakening of the expanding conduction front, note that, at the
end of the heating, the temperature is below 10 MK farther than $2
\times 10^9$ cm from the center of the heating release and below 4 MK
for $R > 10^{10}$ cm.

As soon as the chromosphere is hit and heated by the conduction front,
it expands upwards with a shocked density front\footnote{The code does
not include the effect of plasma viscosity, which has been shown
to be important in solar flares and may spread out and therefore
inhibit the shock front.}. This is the chromospheric evaporation
modeled in confined solar coronal flares (e.g. Nagai 1980, Peres et
al. 1982). The plasma maximum speed of the evaporation front rapidly
increases to $\sim 600$ km/s in about 30 s and then remains between
600 and 700 km/s. The pressure of the evaporation front along the $Z$
axis increases by another factor 5, reaching a value $\sim 10$ dyne
cm$^{-2}$ at the base of the corona. The density front reaches the
central position of the heating along the $Z$ axis in $\sim 30$~s, and
a height $Z \sim 10^{10}$ cm at the end of the heating (t = 150~s),
corresponding to an average propagation speed of $\sim 600$ km/s.  As a
consequence the density in the central post-shock region increases
above a factor 3 but practically in all the coronal region it is 
$< 10^{9}$ cm$^{-3}$ at any time.

Given the highly localized heating release in a non confined
atmosphere, the evaporation front, which mostly determines the soft
X-ray burst, is not plane-parallel: it is practically semicircular,
stronger along the $Z$ axis and weaker and weaker moving farther and
farther from $R=0$.  The reason is that the conduction front arrives at
the chromosphere first along the central $Z$ axis, and progressively
later at greater and greater distances from it. The front amplitude
gets weaker and weaker and with a smaller pitch angle. For $R >
10^{10}$ cm the evaporation speed is $< 100$ km/s and the local density
enhancement very small at any time.

The density in the post-shock region is not uniform: the plasma behind
the shock moves faster than at the shock and accumulates there; the
evaporation front becomes then a bow front and a region of low density
(a proper depression) forms behind it. The bow front fades farther and
farther from the central head.  The diameter of the front is less than
$2 \times 10^{10}$ cm during the whole heating phase.

\subsubsection{The decay phase}
\label{sec:decay}

Fig.\ref{fig:hydro2} shows distributions of temperature, density,
pressure and vertical component of velocity along the central Z axis at
various times of the decay phase, in the reference simulation.

As the heating is switched off (t = 150 s), the temperature suddenly
decreases in the heating region: it halves (from 12 to 6 MK) in about
5~s at the center of the heating release. The cause is the cooling by
thermal conduction, particularly efficient because the density of the
hot plasma remains relatively low ($< 10^9$ cm$^{-3}$). The
cooling front propagates radially from the heating region, while the
internal temperature gets lower and lower. The hot front produced by
the heating impulse has instead weakened and practically disappeared
when reaching the higher and relatively hotter (but very thin) corona.
After one minute since the heating switch off (t $\approx 210$~s), the
temperature along the $Z$ axis overshoots below the initial
temperature of $\approx 1$~MK. At t~=~300~s only a shell at $\sim
10^{10}$ cm from the system origin, and $\sim 10^{10}$~cm thick, is still
(slightly) hotter than the initial atmosphere, while the originally
heated region is practically all cooler than 1~MK.

While such a fast cooling is occurring in the heated region, the
evaporation front continues to propagate upwards and outwards. The
shell expands but maintains more or less the same shape, becoming
geometrically thicker and thicker. The head front is at speed around
400 km/s, the back front is steadily slightly below 800 km/s. The
difference in speed causes the accumulation.

The depression formed behind the head front deepens and expands
continuously behind the evaporation front after the heating is
switched off. At first it is confined to a rather small region behind
the head front, but it extends for $\sim 10^{10}$ cm below the
head front at t = 300 s.

At this time, the thickness of the head front, with a density $\sim 5$
times higher than the background atmosphere, is around $5 \times 10^{9}$ 
cm; the front has reached a height $Z \sim 2 \times 10^{10}$ cm above 
the stellar surface and has a diameter (distance from the $R=0$ axis) of 
$\sim 1.5 \times 10^{10}$ cm. The density in the head front reduces from 
$\sim 3 \times 10^{8}$ cm$^{-3}$ to $\sim 10^{8}$ cm$^{-3}$ as it moves 
away, while the region behind gets less and less dense, the
core of it decreasing below $10^7$ cm$^{-3}$. The pressure in that core, but 
even below it, in a layer $10^{10}$ cm thick, is around 0.01 dyne cm$^{-2}$, 
1/10 smaller than the initial one.

After t = 300 s the cooling of the central regions and the expansion of
the evaporation shell progress with no new feature to remark. At
t~=~800~s no temperature enhancement is visible anywhere, while the
evaporation front is still well evident with the top above $3 \times
10^{10}$ cm from the surface, a height where the density of the
background corona is below $10^7$ cm$^{-3}$, and the tail all above
$10^{10}$ cm. The diameter of the shell is $\sim 5 \times 10^{10}$ cm,
its maximum thickness $\sim 1.5 \times 10^{10}$ cm, but the density is
not uniform inside, ranging between $3 \times 10^7$ cm$^{-3}$ (ahead)
and $10^7$ cm$^{-3}$ (behind). The speed of the head front is
decreasing to 200 km/s. The gap behind the front appears to be
gradually filling and reducing.

\subsection{The X-ray emission of the reference flare}
\label{sec:emiss}

From the model results we have synthesized the X-ray emission both in
the ASCA/SIS band and in two relevant X-ray lines, namely Mg XI at
9.169~\AA~and Fe XXI at 128.752~\AA~with peak formation temperatures of
6.5 MK and 10.5 MK, respectively. These two lines are representative of
strong emission lines typically visible in the band of grating
detectors such as those on board Chandra and XMM-Newton and with
formation temperatures relevant to become significant lines during the
reference model. In the longer wavelength line the spectrometers
typically have higher spectral resolution.

The X-ray spectra at the focal plane of the ASCA/SIS and in the two
selected lines, have been computed in the whole computational domain
and at each output time, according to:

\begin{equation}
{\cal E}_{es}(E,r,z,t) = n^2(r,z,t) ~ G(E,T(r,z,t)) 
\label{eq:emis_est}
\end{equation}
\noindent
where $G(E,T)$ is the emissivity per unit emission measure and per unit photon
energy in the case of the ASCA/SIS, and per unit \AA ngstrom in the case of
the spectral lines. The emissivity in the ASCA/SIS band from 0.6 to 14
keV has been obtained by folding the ASCA/SIS effective area and
response with Raymond \& Smith (1977) spectra of isothermal plasma,
assuming full Anders \& Grevesse metal abundances and a small
interstellar medium absorption ($N_H = 10^{18}$ cm$^{-2}$). The
emissivity in the two spectral lines have been extracted from the
CHIANTI database (version 3.02, Dere et al. 1997), assuming ionization
equilibrium computed according to Arnaud and Raymond (1992) and solar
metal abundances (Grevesse and Anders 1991) and using the default
density of $10^{10}$ cm$^{-3}$.

\begin{figure*}
\centerline{\psfig{figure=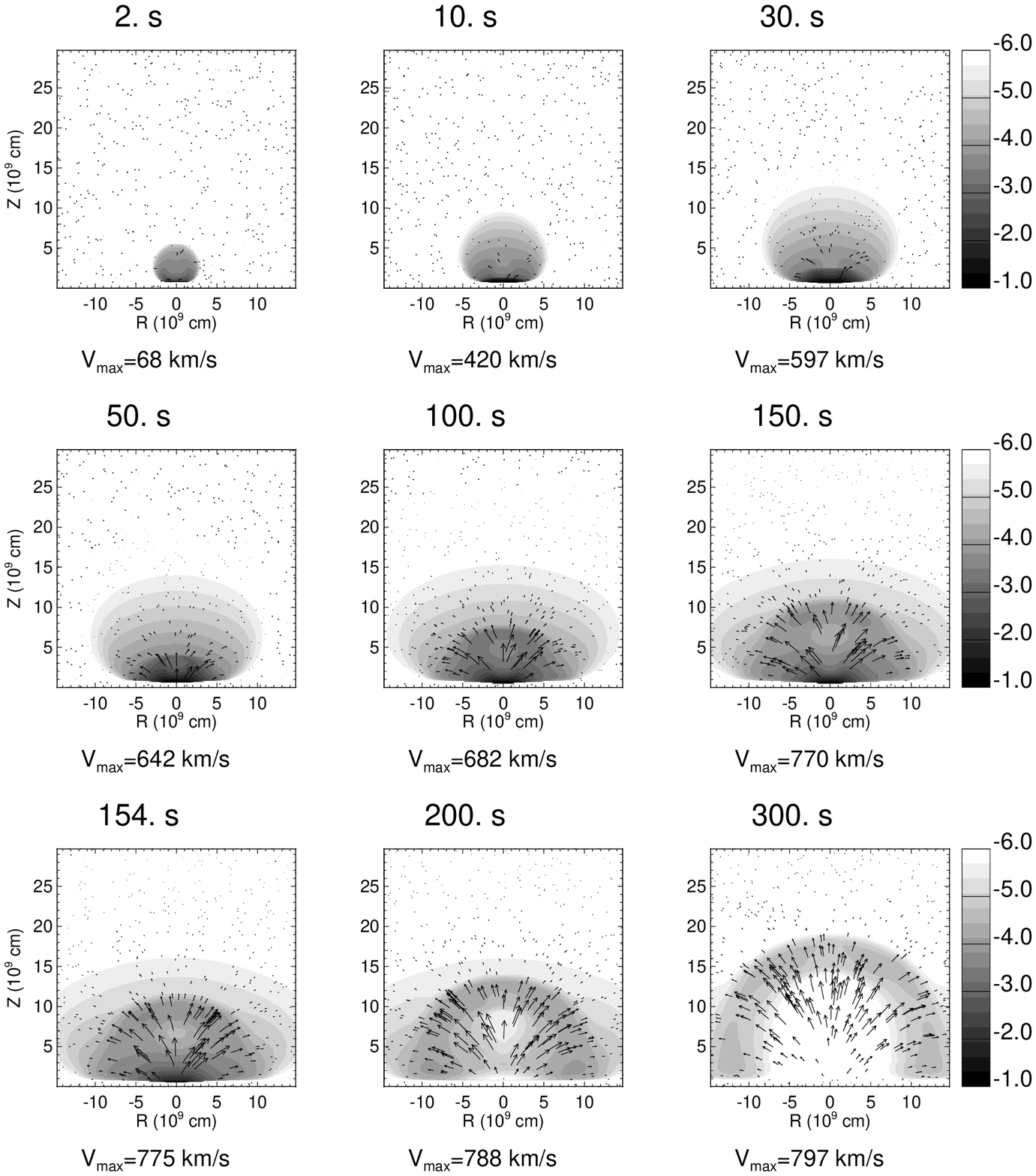,width=14cm}}
\caption[]{Cross-sections (grey scale) of the X-ray emission
(logarithmic scale normalized to global maximum) distribution in the
ASCA/SIS band on a plane across the Z-axis and in the region where the
heating is released, and velocity field ({\it arrows}) at various times
during the reference simulation.  Region, grey scale and maximum speed
as in Fig~\ref{fig:imtemp}.
\label{fig:imasca}}
\end{figure*}

The spatial distribution of the total focal-plane emission per unit
volume at each time is then obtained by integrating in energy:

\begin{equation}
{\cal E}_s(r,z,t) = 
\int_{E_{min}}^{E_{max}} {\cal E}_{es}(E,r,z,t) ~ dE ~~~~~.
\label{eq:emiss_st}
\end{equation}

Fig.~\ref{fig:imasca} shows, for the reference simulation,
cross-sections of the resulting distributions of plasma emission in the
ASCA/SIS band at the same times and in the same region as
Figs.\ref{fig:imtemp} and \ref{fig:imdens}.  The cross-sections are
taken on a plane across the Z axis. The emission at early times is
clearly concentrated in the heating region and mostly in a thin layer
at the base of corona. Until t = 30 s an almost spherical weak emission
front expands progressively and is clearly associated to the
propagation of the thermal front (see Fig.\ref{fig:imtemp}). After t =
30 s another brighter front rises from the base of the corona, with a
less regular shape, and is due to the density front
evaporating from the chromosphere (see Fig.\ref{fig:imdens}). Soon
after the heating is switched off (t = 154 s), the bright knot at the
base of the corona rapidly fades, due to the sudden temperature drop
described in Sect.~\ref{sec:decay}. At t = 200 s most of the emission
comes from the evaporating front, which continues to expands but
gradually fades away (t = 300 s).

Since the flare would not be resolved at stellar distances, we have
then integrated the emission distribution over the relevant region of
the computational domain to obtain the light curve at the focal plane of
ASCA/SIS and the flux at Earth in the two selected spectral lines:

\begin{equation}
{\cal E}(t) = \int_V {\cal E}_s (r,z,t)~ dV
\label{eq:emiss_t}
\end{equation}
where $V=\pi R_{max}^2 Z_{max}$ is the domain volume. In deriving the
focal-plane light curve ${\cal E}(t)$ we have assumed a standard
distance of the flaring source of 1 pc from the detector.

\begin{figure*}
\centerline{\psfig{figure=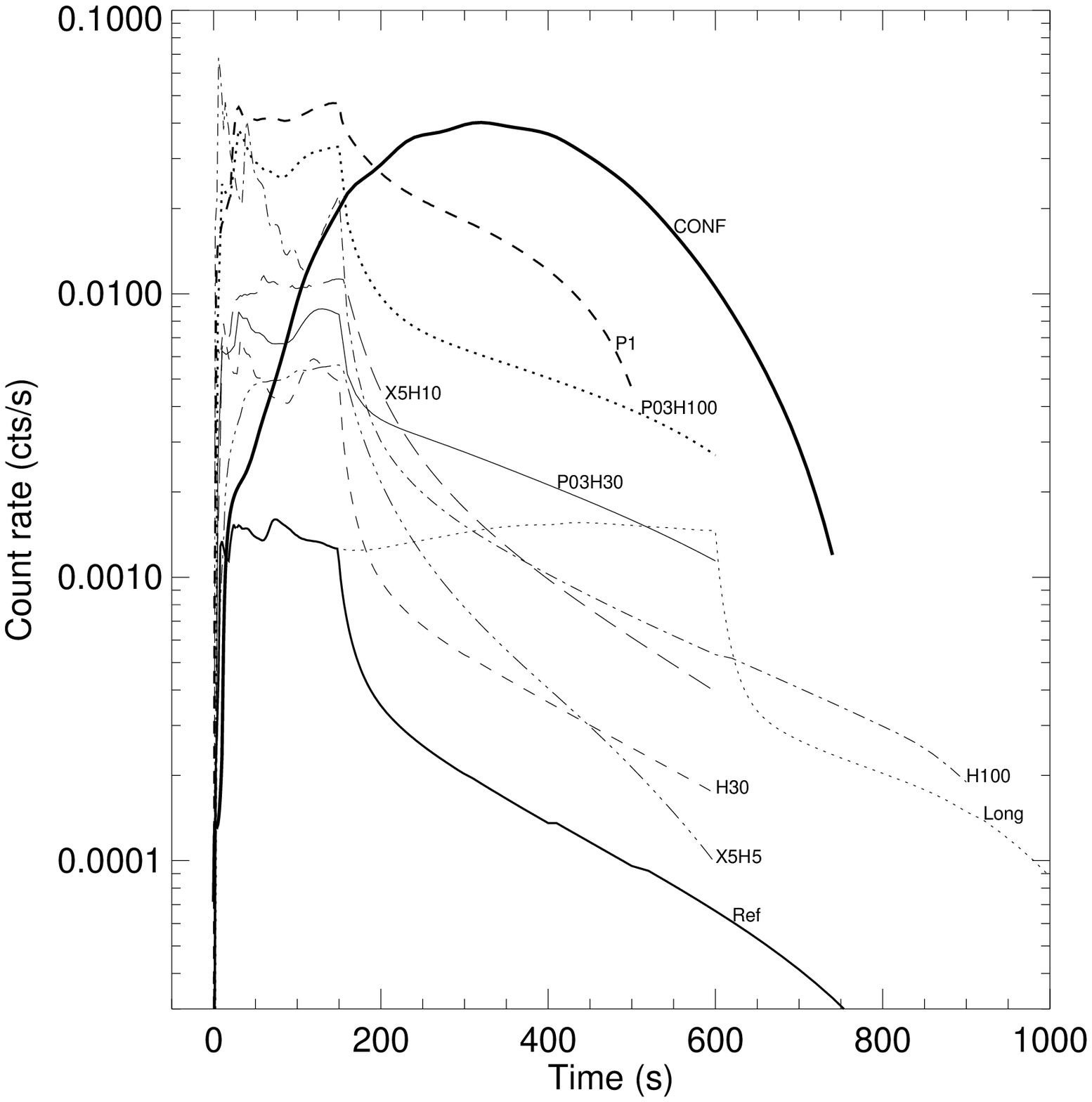,width=14cm}}
\caption[]{Light curves at the ASCA/SIS focal
plane, assuming a distance of 1 pc, synthesized from the results of the
simulations listed in Table~\ref{tab:sim}. The light curve of a
modelled confined flare is also shown for comparison ({\it thickest
solid line}, labelled CONF).
\label{fig:lc}}
\end{figure*}

Figure~\ref{fig:lc} shows the flare net light curves at the ASCA/SIS
focal plane, obtained by subtracting the count rate at the initial time
of each simulation, taken as background non-flaring emission, for all
cases reported in Table~\ref{tab:sim}, including the model of confined
flare. 

Let's first focus the attention to the light curve of the reference
case (the lowest one) and compare it with the confined one: it looks
quite different. The light curve of the confined flare is characterized by a
relatively rapid rise phase, a well-defined peak and a gradual decay. The light
curve of the modeled unconfined flare, instead, after a very rapid rise
($\sim 20$ s) to the maximum value ($\sim 5$ times the initial emission
level), shows a long plateau, staying virtually constant, with small
fluctuations ($\sim 20$\%), until the end of the heating (150 s). As
soon as the heating is switched off, the light curve suddenly drops:
the emission halves in less than one minute. We will call this phase
{\it Fast decay}.  Since then on, the emission decreases more gradually
with an e-folding decay time of $\sim 5$ min, as reported in
Table~\ref{tab:sim}. We will call this phase
{\it Gradual decay}. We notice that the light curve of the non-confined
flare follows much more closely the evolution of the heating: in
particular, the emission drops as soon as the heating stops whereas, in
the confined case, it keeps on increasing and begins to decay only
$\sim 200$ s after the end of the heating. The decay trends are
complementary: in the confined case it is slow at first and then
gradually becomes faster and faster, also because more and more plasma
cools and is no longer visible in the instrument passband.

We have made the exercise of a $\chi^2$ fit of ASCA/SIS spectra sampled
along the light curve with isothermal plasma spectra, a standard
analysis in stellar coronal physics. We obtain a sequence of
temperature and normalization (emission measure) values which can be
put in a density-temperature diagram (where the square root of the
emission measure is used as proxy of the density, as it has been
typically done for confined flares\footnote{Notice that for
non-confined flaring plasma, this assumption, based on constant volume,
is grossly incorrect.}, e.g. Sylwester et al. 1993, Reale et al. 1997), in
order to compare this work with previous work to model stellar flare decay
(Reale et al. 1997, Reale \& Micela 1998, Favata \& Schmitt 1999).
This diagram is used as a diagnostics of the possible presence of
sustained heating during the decay (Sylwester et al. 1993).

\begin{figure}
\centerline{\psfig{figure=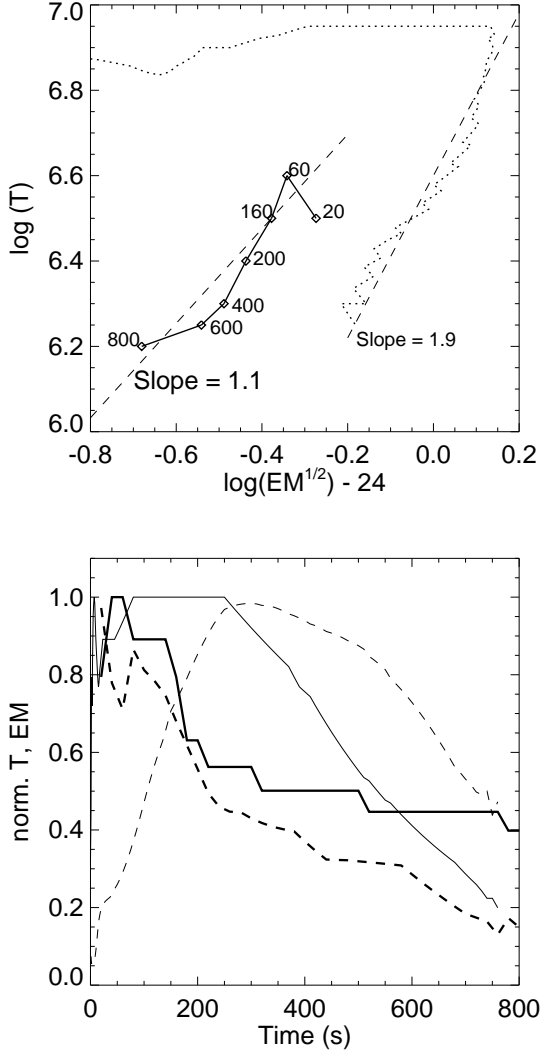,width=8cm}}
\caption[]{{\it Upper panel}: Path in the Density-Temperature diagram 
obtained by fitting the ASCA/SIS
spectra synthesized from the reference model results with isothermal plasma
models at various times (s) during the computed evolution ({\it solid line}). 
The {\it dashed line}
has the labeled slope and well describes the path of the flare decay.
The path of the confined flare model is also shown for
comparison ({\it dotted line}).
{\it Lower panel}: evolution of the temperature ({\it solid lines}) and
of the emission measure ({\it dashed lines}) of the
reference model ({\it thick lines}) and of the model of confined flare ({\it
thin lines}). For the sake of clarity, all quantities are normalized to 1:
the normalization factors are 4.0 MK and 8.9 MK for temperature,
respectively, and $0.3 \times 10^{48}$ cm$^{-3}$ and $1.4 \times 10^{48}$
cm$^{-3}$ for emission measure.
\label{fig:nt}}
\end{figure}

The upper panel of Fig.~\ref{fig:nt} shows the path of the reference
non-confined flare in the density-temperature diagram, and, for
comparison, the one of the confined flare. We first notice that the
maximum best fit temperature of the non-confined flare 
($\sim 4$ MK) is much lower than the true maximum
temperature (see Table~\ref{tab:sim}), by more than a factor 2.  The
reason is that the best-fit temperature is a weighted average over the
emitting region, which is much larger and cooler than the directly
heated region; in the latter the temperature is close to the plasma maximum
temperature.  We end up with an ``observed" temperature well below
typical flare temperatures, although there is plasma above 10 MK and
the rate of energy released is enough to produce a medium (M GOES
class) solar flare, as the confined one shown in the figure, with peak
plasma temperatures around 20 MK.  The emission measure values of the
reference model are lower, by a factor 5, than those of the confined
model flare, with a maximum value of $\sim 3 \times 10^{47}$ cm$^{-3}$
versus $\sim 1.6 \times 10^{48}$ cm$^{-3}$.

The flare path in the diagram is different from the one of the confined
flare in the heating phase: the emission measure reaches very
soon its maximum and then gradually decreases (leftwards), and the
temperature slightly fluctuates, whereas the model of confined
flare shows increasing emission measure at constant temperature
(Jakimiec at al. 1992), as also observed in many solar flares
(Sylwester et al. 1993). The path is much more "standard" in the decay
phase:  temperature and emission measure both decrease, along the line
with positive slope, very similar to many observed cases. The value of
the slope of the non-confined case would suggest a moderately heated
decay (e.g. Reale et al.  1997, Favata et al. 2000) if interpreted as a
confined flare. At the end of the decay the average temperature stays
relatively long at $\sim 2$ MK.

The unusual trend of the n-T path in the heating phase is directly
connected to the relative timing of the evolution of the density and
the temperature (shown in the lower panel of Fig.~\ref{fig:nt}).  In
confined flares, the temperature reaches its maximum and begins to
decay well earlier than the emission measure; the time delay is $\sim
200$ s in the case shown. This does not occur in the reference model of
non-confined flare (as well as in all the other models in
Table~\ref{tab:sim}): the emission measure is at its maximum very soon,
simultaneously to the temperature, and its evolution is 
similar to the temperature evolution and on similar 
time scales.

\begin{figure*}
\centerline{\psfig{figure=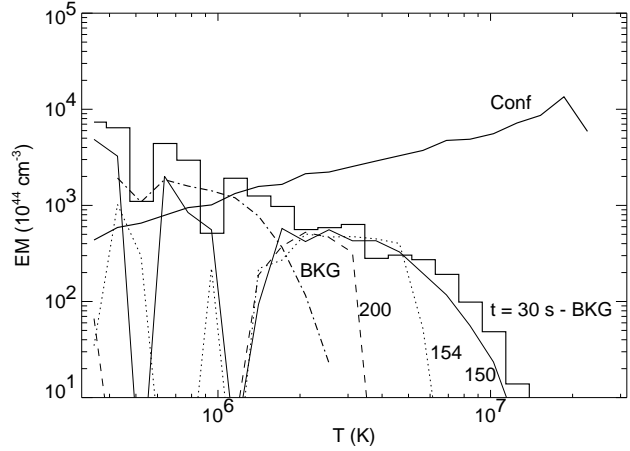,width=9cm}}
\caption[]{Distributions of emission measure vs temperature of the
reference model at various times during the flare evolution: in the
heating phase 30 s ({\it histogram}), 150 s ({\it solid line}); in the
decay phase 154 s ({\it dotted line}), 200 s ({\it dashed line}). The
distributions shown are subtracted by the initial unperturbed
distribution at time (t = 0 s), also shown for comparison ({\it
dashed-dotted thick line}). A distribution obtained from the model of
confined flare (at the end of the heating phase, t = 150 s) is also shown for
comparison ({\it thick solid line}).
\label{fig:emt}}
\end{figure*}

In addition to the different timing of the temperature and emission
measure evolution, also the distributions of the emission
measure versus temperature
during the evolution of the non-confined model flare shows
remarkable differences from corresponding ones during the evolution of
the confined model flare.  Fig.~\ref{fig:emt} shows some distributions
of the emission measure versus temperature (EM(T)) sampled at various
times during the reference model of non-confined flare as compared to
one EM(T) obtained from the confined flare model at the end of the
heating phase (t~=~150~s). In order to evaluate the effect of the flare
heating on the plasma, the figure shows the EM(T) profiles of the
reference model subtracted of the unperturbed pre-flare EM(T) profile.
In agreement to Fig.~\ref{fig:nt}, the maximum amount of emission
measure is present during the heating phase (see profiles at t = 30 s
and t = 150 s). The effect of the heating is to add emission measure at
temperatures above 2 MK, up to $\sim 10$ MK.  Interestingly, the
heating produces additional, and even larger, components of emission
measure at lower temperatures. The overall EM(T) profile in the heating
phase of the reference model therefore {\it decreases} with
temperature, according, approximately, to a power law of index $\sim
-1.5$. This trend is qualitatively different from the typical EM(T)
profiles of flaring (and non-flaring) loops, which instead typically
{\it increase} with temperature with an index $\sim 1.5$.  As the
heating is switched off, the hottest part of the EM(T) suddenly
disappears (t = 154 s), and at 200 s there are no pre-flare components
hotter than 4 MK.  Furthermore, gaps of emission measure form for $T
\leq 1$ MK: the plasma in the heated region cools and only the plasma
in the outward expanding shell contributes to the emission measure in
excess of the background.

\begin{figure*}
\centerline{\psfig{figure=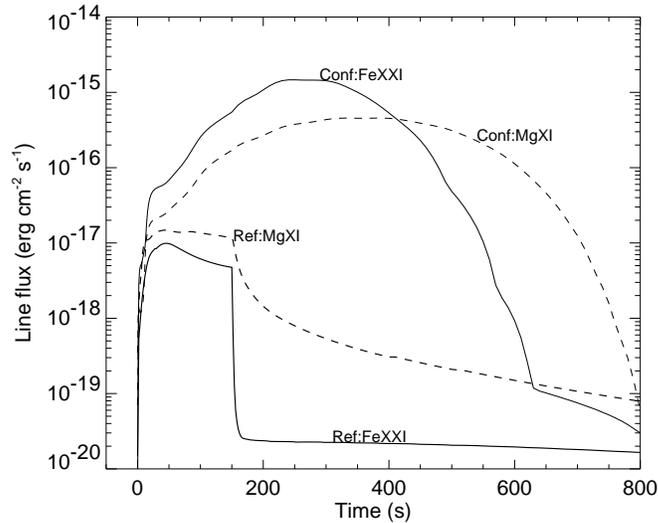,width=9cm}}
\caption[]{Light curves in the two selected X-ray lines for the reference model
and for the confined flare model.
\label{fig:lc2}}
\end{figure*}

As shown in Fig.~\ref{fig:emt}, emission measure components are present
in a wide range of temperatures during the heating phase and, therefore, we can
expect significant emission in the selected spectral lines.
Fig.~\ref{fig:lc2} shows the light curves of the reference non-confined
flare in the selected Mg XI at 9.169~\AA~and Fe XXI at 128.752~\AA~
lines, assumed always in equilibrium of ionization. The corresponding
light curves for the confined model flare are also shown for
comparison. The light curves of the reference flare globally resemble
the other synthesized ones at the ASCA/SIS focal plane
(Fig.~\ref{fig:lc}). Their decay starts simultaneously. While the light
curve in the hotter Fe XXI line drops by more than two orders of
magnitude in less than 5 s, due to the sudden cooling of the heated
region, the Mg XI line decays more slowly, because emission measure
survives longer at temperatures below 10 MK.  Once again, the light
curves look quite different from the corresponding ones of the confined
flare model. In particular, the cooler Mg XI line peaks later than the
hotter Fe XXI line, which is more directly sensitive to the evolution
of the heating.  The shapes of both the line decays for the reference flare
are different from those of the confined flare, but it may not be
trivial to discriminate between them in real data, typically coarsely
resolved in time, and sometimes with small signal to noise ratio.

\begin{figure*}
\centerline{\psfig{figure=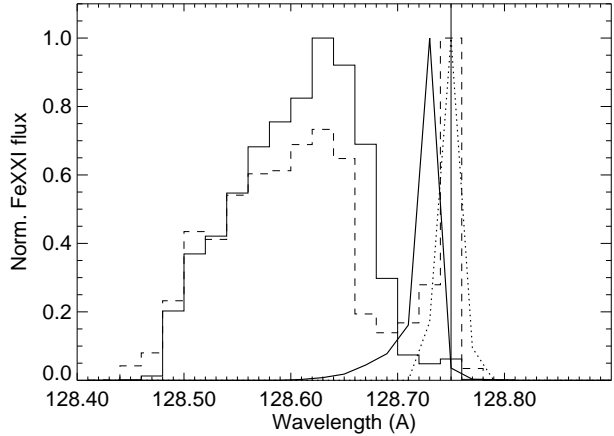,width=9cm}}
\caption[]{Fe XXI 128.75 \AA~ line profiles synthesized from the
reference model and averaged over the heating phase (from 0 to 150 s,
{\it solid histogram}), and over the next 150 s in the decay phase
({\it dashed histogram}).  Analogous profiles from the confined model,
averaged over the heating phase (from 0 to 150 s, solid line) and over
the next 150 s ({\it dashed line}) are also shown for comparison. The
normalization factors are $3.9 \times 10^{-19}$, $6.0 \times 10^{-22}$,
$8.0 \times 10^{-17}$, $2.9 \times 10^{-16}$ erg cm$^{-2}$ s$^{-1}$
\AA$^{-1}$, respectively. The spectral bin is 0.02~\AA. The nominal line
wavelength is marked ({\it vertical line}).
\label{fig:lspec}}
\end{figure*}

As discussed in section~\ref{sec:evol}, the evolution of the non-confined plasma
has showed the presence of a steady upward moving plasma front, with speed
of several hundreds km/s. If favorably oriented with respect to the
line of sight, we may expect such high speed plasma component to produce
significant Doppler blue-shifted components in resolved lines.
Fig.~\ref{fig:lspec} shows two detailed profiles of the
Fe XXI 128.75 \AA~ line synthesized from the reference model, in the
assumption that the flare is "disk-centered" with respect to the
observer. One profile is averaged over the heating phase (from 0 to 150
s), and the other over the next 150 s in the decay phase.  The former
profile is practically entirely blue-shifted, while the other also
shows a significant static component. The blue-shifted components both
peak at 128.63 \AA, corresponding to an outward speed component of
$\sim 300$ km/s, and are considerably broadened, with a width of
0.05 -- 0.10 \AA~ (100 -- 200 km/s).  Such persistent
blue-shifted components are not present in the profiles from the model
of confined flare.  The profile averaged over the whole heating phase
is slightly blue-shifted (0.02~\AA, 50~km/s), due to the plasma
evaporation, the other profile is practically unshifted. Both profiles
are much sharper than those of the reference model. Indeed considerable
blue-shifted components (at speed $\geq 100$ km/s) are predicted also
by models of confined flares, but only in the very initial phase of the
plasma evaporation, generally within the first 60 s in an active region
loop. Then the blue-shift is washed out by the next more gradual, 
less dynamic (but much brighter) evaporation.

\subsection{The other simulations}

As discussed in Sec.~\ref{sec:parref}, the basic free parameters of the
simulations are the pressure at the base of the corona and the heating
parameters. The explored values are shown in Table~\ref{tab:sim}. We
have already described in detail the results of the reference
simulation. The plasma evolution in all the other simulations
invariably resembles the one obtained in the reference case, i.e.
initial temperature increase and expanding conduction front, subsequent
bow density front expanding upwards from the chromosphere, decay with
very fast temperature decrease and continuous expansion of the density
bow front. However several quantitative differences occur. First of
all, the maximum temperature is sensitive to the heating parameters
and, in particular, it increases with the rate per unit volume $H_0$
and with the height $Z_H$. Given the high efficiency of thermal
conduction, the temperature in a flare reaches an equilibrium value
very rapidly. In steady-state closed coronal structures (loops) the
plasma maximum temperature is linked to the heating amount by coronal
scaling laws (Rosner et al. 1978, hereafter RTV). One may then use the
temperature of a loop to estimate the heating rate with the aid of the
scaling laws (e.g. Reale et al.  2000, Maggio et al. 2000). In the
presence of the same physical effects at work, we have applied the
scaling laws also to our configuration despite the scaling laws were
devised for closed one-dimensional systems.

By combining the maximum temperature scaling law to the volume heating
scaling law we obtain:

\begin{equation}
L \approx 10^{-3} T^{7/4} H^{-1/2}
\label{eq:rtvn}
\end{equation}
where $L$ was the loop half length in the RTV scaling laws, $T$ is the
plasma maximum temperature and $H$ is heating volume rate. If we assume
$T = T_{max}$ and $H = H_0$, we invariably obtain $L \sim 10^9$ cm for
the cases named Ref, H30 and H100 in Table~\ref{tab:sim} and $L \sim 2
\times 10^9$ cm for both the cases X5H5 and X5H10. From
Table~\ref{tab:sim}, we realize that these lengths correspond
essentially to the ``diameter'' of the heating distributions, i.e.
twice the width $\sigma_H$. This means that in this non-confined
configuration the thermal distribution of the flaring region is
virtually dominated by the heating distribution.

The maximum temperatures of the simulations range between 13 and 25 MK.  Notice
that, in order to produce a 25 MK flare, not indeed a major one on the
standard of stellar flares, we have needed quite a high heating rate of
$2 \times 10^{29}$ erg/s.  We notice also a weak inverse dependence of
the maximum temperature on the coronal initial base pressure: the heating is
less effective in an atmosphere initially at higher pressure.

Fig.\ref{fig:lc} includes light curves at the ASCA/SIS focal plane for all 
simulations in Table \ref{tab:sim}, all computed as described in 
Sect.~\ref{sec:emiss}. Looking at the figure, the most striking feature is
certainly that, apart from a scale factor clearly depending on the
amount of energy released in the atmosphere, all light curves share a
common shape: a very fast rise, a relatively constant phase, lasting
until the heating is switched off, a sudden fast decay by a factor 2 to
5 on time scale of few seconds, followed by a more gradual decay on
time scales of few hundreds of seconds. 
This evolution and its timing traces closely the evolution of the
heating.  

It is straightforward to attribute the fast initial cooling to thermal
conduction. From Eq.\ref{eq:energy}, the conductive cooling time can be
approximated as:

\begin{equation}
\tau_c \sim 5 \frac{n_9 l_9^2}{T^{5/2}_7} ~~~ \rm s
\label{eq:tauc}
\end{equation}
where $n_9$ is the density in units of $10^9$ cm$^{-3}$, $l_9$ is a 
characteristic length ($10^9$ cm) and $T_7$ is the peak temperature 
($10^7$ K). In practice, the relatively small density value makes the 
cooling by conduction particularly efficient. As shown by Table 
\ref{tab:sim}, the low density is common to all simulations and is an
intrinsic, important and characterizing feature of 
simulating non-confined flaring plasma, as further discussed below.
Notice also that the highest pressure case (P1) shows a relatively
higher density and, in fact, a relatively more gradual initial decay.

Later on, as temperature decreases, cooling by conduction becomes less
and less efficient and other cooling mechanisms come into play,
determining the flattening of the light curves in the gradual decay
phase. In particular, in the analysis of the simulation results
(Fig.\ref{fig:imasca}) we have noticed that in the gradual decay phase
the flare emission originates mostly from the expanding density shell.
The front propagates in a medium reasonably approximable as isothermal,
and, therefore, the gradual decay must be attributed to the progressive
reduction of the emission measure. Since, as we have checked, the
amount of plasma involved in the density front is approximately (within
$\sim 20$ \%) unchanged with time as the expansion progresses, the
emission measure in this phase is roughly proportional to density. As
the front propagates upwards, it maintains approximately the same
density contrast, as it occurs in strong shocks.  Therefore the density
of the front decreases simply because it moves upper and upper through
less and less dense layers of the stratified corona. The density
decreases exponentially:

\begin{equation}
n \approx n_0 \exp(z/s_p) \approx n_0 \exp(-t/\tau_f)
\end{equation}
where $s_p \approx 2 k T_f / (m g) \sim 5000 ~ T_f$
is the pressure scale height of the front at temperature $T_f$
for solar gravity,
and 

\begin{equation}
\tau_f = s_p/V_f \approx 300 \frac{T_{6.5}}{V_{500}}
\label{eq:tauf}
\end{equation}
is an e-folding time obtained using the fact that the front moves at
approximately constant speed ($V_f \sim 500$ km/s).  The time $\tau_f
\sim 300$ s, obtained for $T_f \sim 3$ MK is indeed of the order of the
decay times reported in last column of Table \ref{tab:sim}.

Of course, the decaying trend changes again
as soon as any of the above approximations no longer holds, e.g. as
soon as the shock weakens (as in case P1 for $t \ga 400$ s) or is very
distant from the origin (as in the case named {\it Long} for $t \ga 900$ s).

All light curves share the same fundamental differences from the light
curve of the confined model flare: in particular the decay starts much
earlier and starts fast and then is more gradual, whereas in the
confined flare it starts slow and then steepens. It should be noted
that, while the only way to have a significantly slower flare decay in
the non-confined atmosphere is to have a heating released over a much
larger area (Eq.~\ref{eq:tauc}), in contrast with flare heating theory
(e.g. Golub \& Pasachoff 1997), the decay of confined flare can be made
slower simply considering a longer loop, according to the scaling law
(Serio et al. 1991):

\begin{equation}
\tau_{conf} \approx 120 \frac{L_9}{\sqrt{T_7}}
\label{eq:tauserio}
\end{equation}

\section{Discussion}
\label{sec:discuss}

Modeling a non-confined coronal flare has produced interesting results
from the points of view both of pure modeling and of diagnostics,
especially when compared to confined events.

\begin{table*}
\begin{center}
\caption[]{\label{tab:disc} Relevant properties of flare models$^a$}
\begin{tabular}{lp{4.cm}p{4.cm}p{4.cm}p{3.cm}}
\hline
Item&Property&Confined&Non-confined&Diagnostic power: tools$^b$\\
\hline
\multicolumn{3}{l}{Hydrodynamics}&&\\
\hline
1&Flaring region&All loop&Localized: heated region& -- \\
2&Scale length in RTV scal. laws&Loop half-length&Diameter of heated
region& --\\
3&Evaporation evolution&Plasma accumulation&Expanding fronts& --\\
4&Decay time scale&$120 L_9/\sqrt{T_7}$&early: $5 n_9 l_9^2/T_7^{7/2}$&--\\
&&&late: $300 \sqrt{T_{6.5}}/V_{500}$\\
\hline
\multicolumn{3}{l}{Hydrodynamics \& Diagnostics}&&\\
\hline
5&Flare temperature&High$^c$&Lower&Low: LRS$^{d}$\\
6&Plasma density&High ($\ga 10^{11}$ cm$^{-3}$)&Low ($\la 10^{10}$ cm$^{-3}$)& Low: HRS$^{e}$\\
7&T-EM relative timing&Peak delay&Synchronous&High: LRS\\
8&Decay&Gradual&Fast$+$Gradual&Very high: WLC$^{f}$\\
9&Line evolution&Asynchronous&Synchronous&?: HRS\\
10&Blue-shift phase&Short&Persistent&Low: Very HRS $+$ orientation$^g$\\
11&EM(T)&$\propto T^{3/2}$&$\propto T^{-3/2}$&?: MRS$^{h}$\\
12&Heating tracer&Hard X-ray l.c.&Soft X-ray l.c.&?: WLC\\
\hline
\end{tabular}
\end{center}
\noindent
$^a$ This table summarizes the main differences between the results of
confined and non-confined flare models, discussed in more detail in
Sect.~\ref{sec:discuss}

$^b$ Level of diagnostic power to discriminate confined and non-confined
flare: Low, High, Very high. "?" means uncertain 

\noindent
$^c$ With the same heating rate: the value depends on the heating rate.

\noindent
$^d$ LRS: Low Resolution Spectra required

\noindent
$^e$ HRS: High Resolution Spectra required

\noindent
$^f$ WLC: Wide-band Light Curve required

\noindent
$^g$ the flaring region must be located well inside the stellar disk

\noindent
$^h$ MRS: Medium Resolution Spectra required

\end{table*}

We have considered a reference case of a flare triggered in an open
structure by a heating pulse which would have produced a 20 MK flare in
a confined structure, namely a 40,000 km long solar loop. Then we have
explored several other cases, by increasing the heating rate, therefore
approching more typical stellar flare conditions, by shifting upwards
the heating location, as it may occur in spatially large stellar
coronae, and by considering initial atmosphere at higher pressure,
which may be present in active stars.

Table~\ref{tab:disc} shows the main results of the present study, by
comparing properties of non-confined flare models with those of
confined flare models and indicating the potential way of
discriminating them through the analysis of observations.

The hydrodynamic evolution obtained from modeling non-confined flares
is complex and involves many different aspects. The model flare is
triggered by an impulsive heating in localized region of a stratified
corona. This heating produces an immediate but localized temperature
increase: proper hot flare conditions occur only within a relatively
small region around the heating deposition ({\it Item 1 in 
Table~\ref{tab:disc}}). This is at variance from a
flare occurring in confined plasmas, which are typically entirely
involved in the flare. In our case, thermal conduction has a more
limited action range: the conduction front rapidly weakens due to its
expansion and is not able to heat significantly plasma very far from
the heating deposition place.

In spite of the dynamic flare evolution, the plasma reaches soon a
steady state thermal condition, so that the maximum temperature and
heating rate per unit volume are approximately ruled by a scaling law
similar to RTV scaling laws for coronal loops, in which the
heating scale length takes place of the loop half-length
(Eq.~\ref{eq:rtvn}, {\it Item 2 in Table~\ref{tab:disc}}).

A crucial difference from confined flare events is the evolution
of the chromospheric evaporation ({\it Item 3 in
Table~\ref{tab:disc}}). In a closed structure, plasma evaporates first
rapidly with a strong density front, and then more gradually but {\it
continuously} until the loop is completely filled up and is close to
equilibrium conditions. This long-lasting evaporation in a confined
structure allows density to reach high values ($\ga 10^{11}$
cm$^{-3}$). The high density (together with the flat temperature
distribution along the loop) makes cooling by conduction relatively
inefficient, leading to decay times of hours in long loops.  What
happens in the non-confined cases is radically different.  There is a
single evaporation front, which propagates as a {\it strong shock}
continuously outwards from the chromosphere below the heating place,
but since plasma is not confined, it does not accumulate in spite of
further evaporation from the chromosphere. In other words, even though
the density in the flaring region soon becomes several times higher
than it was in the unperturbed condition, it does not grow above the
strong shock limit (e.g. Rosenau \& Frankenthal 1976), and {\it it
never becomes as high as in typical confined flares}.  This is an
unavoidable property of a non-confined configuration:  plasma cannot
pile up and therefore cannot become very dense.  This has several
implications, but probably the most important one, especially from the
diagnostical point of view, is that, as soon as the heating stops, the
plasma in the heated region cools very rapidly, because thermal
conduction is very efficient in a relatively low density medium.  We
have provided a characteristic time scale for this early and fast decay
phase (Eq.~\ref{eq:tauc}), which, for typical conditions, is of the
order of few seconds ({\it Item 4 in Table~\ref{tab:disc}}).

We have seen that, after the initial drop, the decay progresses more
gradually, driven by the decrease of density of the emitting plasma
contained within the expanding evaporation front, with a time scale
given by the time taken by the front to move across the pressure scale
height (Eq.~\ref{eq:tauf}).  However, even the characteristic decay
time of this phase, invariably not longer than few minutes, is quite
small as compared to the decay times of typical observed large stellar
flares (hours or even days), ruled by the conduction/radiation cooling time
(Serio et al. 1991).

The thermal and hydrodynamic evolution of the flaring plasma leads to a
series of diagnostical implications, which can be searched for in
observations of stellar flares. In the first place, we note that
non-confined flare require much more energy to enhance the plasma
temperature ({\it Item 5 in Table~\ref{tab:disc}}): with the same heating rate both the maximum (see
Table~\ref{tab:sim}) and the average (see Fig.~\ref{fig:nt})
temperatures of the non-confined flare are nearly halved with respect
to the corresponding ones of the confined model. As obvious consequence
one may expect long-lasting stellar flares to be also relatively cool,
which is opposite to what is typically observed, with temperature even
reaching $10^8$ K (e.g. Maggio et al. 2000).  However, since
one cannot exclude very high energy rates, the high temperatures
observed can hardly be taken as a distinctive signature of
confinement, in the light of our findings.

As discussed before, the plasma density should be instead invariably
lower in non-confined flares ({\it Item 6 in Table~\ref{tab:disc}}).
This may be directly diagnosed with density-sensitive lines detected
with high resolution spectrometers such as the gratings on board
Chandra or XMM-Newton. However, it is not trivial to obtain such data
for flares and to date no such diagnostics is available yet.

It is instead well-known that in many flares, both solar and stellar,
the temperature peaks and starts to decay well before the emission
measure. This is consistently described by models of confined flares:
as the heating drops, the temperature begins to decay, while plasma
evaporates and makes emission measure increase still for some time.  In
our simulations of non-confined flares, instead temperature and
emission measure both decay immediately and simultaneously after the
heating stops. This difference may indeed be a strong indication of
confinement for long-lasting stellar flares ({\it Item 7 in
Table~\ref{tab:disc}}).

Another significant indication comes from the analysis of the light
curves in a wide X-ray passband, such as ASCA/SIS ({\it Item 8 in
Table~\ref{tab:disc}}). In
the case of non-confined flares the light curve invariably drops as
soon as the heating stops, with a decay time in the early phase of few
seconds (Eq.~\ref{eq:tauc}), and then to decay more gradually, with a
time scale of few minutes (Eq.~\ref{eq:tauf}).  The early decay of the
light curve of the confined flare is much more gradual, and becomes
even more gradual (hours) if the confining loop is longer than a solar
active region loop (Eq.~\ref{eq:tauserio}).  Unless we hypothesize unlikely
and {\it ad hoc}
heating depositions, e.g. decreasing very gradually and
completely driving the decay phase, 
{\it a sudden drop of the X-ray light curve is an
expected intrinsic and unavoidable feature of flare events in a
non-confined corona}.

The light curve in selected X-ray lines are also different when
comparing confined and non-confined events ({\it Item 9 in
Table~\ref{tab:disc}}). In particular, in the model of confined flare
the light curves in two lines at different formation temperature are
considerably asynchronous: the hotter line peaks and decays earlier. In
the non-confined case the two lines evolve practically with the same
timing.  This diagnostic may become interesting if flare light curves
in single X-ray lines are available.

The presence of long-enduring, high speed, expanding plasma fronts may
open the opportunity to detect persistent blue-shifted components in
high resolution line profiles ({\it Item 10 in Table~\ref{tab:disc}}). Upflows of about 400 km/s have
been seen in the Ca XIX and Fe XXV lines in the early thermal phase of
some solar flares (e.g. Antonucci et al. 1987, 1990). Hydrodynamic
models of confined flares predict that the initial chromospheric
evaporation may produce blue-shifts in X-ray lines, but for very short
times (less than one minute in active region loops, see
Fig.~\ref{fig:lspec}).  We have shown that a steady broadened component
blue-shifted by $\sim 300$ km/s may be expected in the FeXXI line at
128.75 \AA~ for the reference non-confined model. This kind of
prediction may not provide significant indications when looking at real
flare observations for various reasons: blue-shifts are expected only
if the flaring region is favourably oriented to the observer; shifts of
few hundreds km/s are hardly detectable even with XMM and Chandra
high
resolution grating spectrometers; although less persistent, significant
blue-shifted components may be present even during confined flares in
long loops.  All these limitations may make any diagnostics of
confinement from line shifts rather difficult.

Recent works have pointed the attention to the derivation of the
distribution of emission measure with temperature ({\it Item 11 in
Table~\ref{tab:disc}})
from the analysis of
X-ray stellar data (e.g. Drake et al.  2000, Linsky \& Gagne 2001).
Non-confined flare models predict a flare emission measure distribution
monotonically decreasing with temperature during the heating
phase and then losing the highest temperature components and some
intermediate temperature components in the decay phase. These
distributions are quite different from those typically derived from
flaring and non-flaring solar (e.g. Peres et al. 2000, Reale et al.
2001) and non-flaring stellar data, which instead show well-defined EM
peaks. This may be a discriminating feature, whenever medium resolution
spectra provide reliable emission measure distributions during stellar
flares (e.g. Osten et al. 2000).

As a final diagnostic consideration, we point out that, if data show
evidence of a flare occurring in a non-confined plasma, then according
to our modeling the soft X-ray light curve would faithfully trace the
evolution of the heating, an important diagnostics for coronal physics
({\it Item 12 in Table~\ref{tab:disc}}).  An equivalent information is
generally derived from hard X-rays in solar flares (e.g. Golub \&
Pasachoff 1997).

We may wonder whether our results, such as the light curves
presented in Fig. \ref{fig:lc}, are affected by the Non Equilibrium of
Ionization (NEI) conditions of the plasma. In general, in fact, the
X-ray emissivity depends on the thermal history of the emitting plasma
parcel, which is unfortunately not straightforward to evaluate. Bocchino
et al. (1997) show that the 0.1-2.0 keV X-ray emissivity of an
impulsively heated plasma (a situation similar to the rise of the flare
in our models) approaches the equilibrium value for ionization time
$\log\tau \sim 11$ in s cm$^{-3}$.  This is also true in the ASCA/SIS
bandwidth.  In the initial phase of the flare the typical density of
the bright X-ray emitting plasma is in excess of $10^9$ cm$^{-3}$,
corresponding to an equilibrium time scale of the order of 100 s: we
may expect a modification of the light curve before 100 s, consisting
in a higher NEI value reached at the end of the rising phase and a slow
descend to the sustained rate.  In the fast decay at the end of the
heating phase the NEI conditions may have some importance and bring to
variations of the light curve in this phase. However, these effects
will be reduced for hotter and hotter events, i.e. typical intense
stellar flares, because the continuum emission dominates more and more
on the line emission. Later on, in the gradual decay, most of the
emission comes from the expanding shock, in which we have only a slight
decrease of the temperature, and therefore the plasma is only
marginally overionized. The small deviation from equilibrium gives
quasi-equilibrium conditions during most of the decay phase, thus we do
not expect substantial modifications of the global time scales we have
worked out.

In summary, there are two basic features that may characterize
invariably flares in non-confined regions and may be relatively easy to
trace in stellar X-ray data: light curves with very fast decays and
synchronous evolution of density and temperature. Both these features
are in contrast with evidence from stellar flares.  Then we come to an
interesting conclusion: {\it the long-lasting X-ray flares typically
detected in several stellar observations cannot occur in non-confined
atmospheres.} This is somewhat unexpected, because long-lasting stellar
flares have been typically associated to very large coronal structures
(see Sect.~\ref{sec:intro}), which have the non-confined
configuration as an asymptotic extreme. This work shows that most
likely the long-lasting flares, such as those in active stars, still
occur in closed, although large, structures and that the role of
confinement of the coronal magnetic field must be invariably
significant in those events. In other words, the magnetic fields must
always be strong and/or the magnetic structure never breaks open.  In
the same direction, notice also that non-confined flares appear to
involve huge amounts of heating (see Table~\ref{tab:sim}) which may not
be entirely realistic when compared to the energy budget of stellar
coronae.

If long-duration stellar flares are unlikely to be described as breaking the
magnetic confinement, are there any other observed events which instead
could be? As mentioned in Sect.~\ref{sec:intro} long stellar flares
are preferentially detected in X-ray observations. However there is a
class of coronal variations which occur on small time scales and which
have been detected on dMe star UV Ceti by ROSAT (Schmitt et al. 1993). The
possibility that such short events may be interpreted as small but very
intense non-confined events should be explored.

This study in some way shows an interesting theoretical perspective to
be compared with SNR models and Coronal Mass Ejections (CME) models.
Our model shows an overheated plasma which expands under the effect of
heating in a hot and thermally conducting stratified atmosphere, and we
provide for it characteristic scalings of general validity
(Eqs.~\ref{eq:rtvn}, \ref{eq:tauc}, \ref{eq:tauf}). For instance, since
the time scale of the late decay $\tau_f$ (Eq.~\ref{eq:tauf}) depends
practically only on the speed of the evaporation front and on the
thermal plasma conditions far from the flaring region, it is little
dependent on the flare heating and on other model parameters: it is a
general decay time scale for shell fronts expanding in a stratified
conducting corona. It should be noted that application to other
astrophysical systems is not trivial.  In SNR models, plasma expands in
a much cooler, thinner and less conducting medium.  On the other hand,
CME's are mostly observed in the UV band, and are therefore to be
modeled as relatively cool perturbations, whereas the non-confined
fronts modeled here are at standard coronal temperatures.  For the
latter fronts to become proper CME's, a mechanism to thermally insulate
the fronts from the surrounding corona should be invoked, so that they
are free to cool by radiation and to emit in the UV band (e.g.
Ciaravella et al. 2001).

\section{Conclusions}
\label{sec:conclu}

Stellar flares are generally observed to evolve and decay on time
scales ranging from several hours to days, suggesting large flaring
regions or even lack of magnetic confinement. This work shows that the
hydrodynamic evolution of flares occurring in non-confined atmospheres
leads invariably to a much faster decay of the brightness (on time
scales of very few minutes) after the heating phase, in a wide range of
the physical parameters. One of the main implications is that the long
duration of stellar flares is not indicative of non-confined plasma.
Reversing the argument, it is highly probable that the observed
long-lasting stellar X-ray flares involve forms of plasma confinement
in closed coronal structures.  We have pointed out other characteristic
features, with more or less diagnostical power, e.g. the synchronous
evolution of density and temperature, summarized in
Table~\ref{tab:disc}, that seem to indicate that X-ray observed stellar
flares involve mostly plasma confined in closed structures.

Although this work simulates rather extreme conditions of flares 
occurring in completely non-confined atmospheres, the result that 
long-lasting stellar flares are likely not occurring in large open 
structures largely motivates its validity. 

Furthermore this work predicts the possible existence of a new
phenomenological class of events, characterized by peculiar light
curves with a limited range of time scales. Such kind of events may be
addressed by observations at high sensitivity levels, which allow for
high time resolution and high signal-to-noise ratio, such as those
obtained from the current missions Chandra and XMM-Newton.

\bigskip
\bigskip
\acknowledgements{This work was supported in part by Agenzia Spaziale
Italiana and by Ministero della Universit\`a e della Ricerca
Scientifica e Tecnologica.  Simulations have been executed on the Cray
T3E at CINECA (Bologna, Italy), on a Compaq server and on a Compaq HPC
Cluster at Osservatorio Astronomico di Palermo (Italy).}

{}

\end{document}